\documentclass[prd,draft,amsfonts,amssymb,amsmath,showpacs]{revtex4}

\begin{document}

\title{Dynamical Chiral Symmetry Breaking and its Restoration for an Accelerated Observer}
\author{Tadafumi Ohsaku}
\affiliation{Yukawa Institute for Theoretical Physics, Kyoto University, 
Kitashirakawa-oiwakecho, Sakyo-ku, Kyoto 606-8502, Japan.}

\date{\today}


\newcommand{\bmx}{\mbox{\boldmath $x$}}
\newcommand{\bmy}{\mbox{\boldmath $y$}}
\newcommand{\bmk}{\mbox{\boldmath $k$}}
\newcommand{\bmp}{\mbox{\boldmath $p$}}
\newcommand{\bmq}{\mbox{\boldmath $q$}}
\newcommand{\bmP}{\mbox{\boldmath $P$}}  
\newcommand{\kfey}{\ooalign{\hfil/\hfil\crcr$k$}}
\newcommand{\pfey}{\ooalign{\hfil/\hfil\crcr$p$}}
\newcommand{\qfey}{\ooalign{\hfil/\hfil\crcr$q$}}
\newcommand{\Deltafey}{\ooalign{\hfil/\hfil\crcr$\Delta$}}
\newcommand{\nablafey}{\ooalign{\hfil/\hfil\crcr$\nabla$}}
\newcommand{\Dfey}{\ooalign{\hfil/\hfil\crcr$D$}}
\newcommand{\partfey}{\ooalign{\hfil/\hfil\crcr$\partial$}}
\def\sech{\mathop{\rm sech}\nolimits}

\begin{abstract}

Based on the Hawking-Unruh thermalization theorem,
we investigate the phenomenon of the dynamical chiral symmetry breaking and its restoration
for a uniformly accelerated observer. 
We employ the Nambu$-$Jona-Lasinio model in Rindler coordinates,
and calculate the effective potential and the gap equation.
The critical coupling and 
the critical acceleration for symmetry restoration are obtained.

\end{abstract}


\maketitle

The Hawking-Unruh effect predicts that 
the accelerated observer sees a thermal excitation of particles of a system in Minkowski spacetime,
with the Unruh temperature $T_{U}=a/2\pi$ ( $a$ is an acceleration constant )~[1].
This is called the thermalization theorem~[2-10].
To study the effect of uniform acceleration, we employ Rindler coordinates
which are appropriate for the spacetime for a uniformly accelerated observer.
It should be noticed that the thermally excited particles are not the original Minkowski particles
but the Rindler particles determined 
by the ground ( vacuum ) state of the system of Rindler coordinates. 
It has been generally proved that, in Rindler coordinates, Euclidean two-point functions are 
periodic in the direction of time with period $2\pi$, and
these functions satisfy the Kubo-Martin-Schwinger condition~[3-7,10].
Therefore, Green's functions in Rindler coordinates with period $2\pi$ 
may be interpreted as finite-temperature Green's functions.

In this letter, we examine what will be observed by a uniformly accelerated observer
who moves relative to a system in which the chiral symmetry was dynamically broken.
The following situation is considered: After a fermion system in Minkowski spacetime 
dynamically generated a chiral mass, an observer will be accelerated uniformly.
The acceleration may give a thermal effect on the observation of the system.
Of particular interest here is the question whether the thermalization effect 
of acceleration gives the restoration of broken symmetry or not.
For the purpose to study the problem,
we employ the Nambu$-$Jona-Lasinio model~[11].
The field theory in Rindler coordinates can be described by the method
of curved background field~[12]. 
We obtain the Green's function in Rindler coordinates, 
and apply the method of effective potential given in Ref. (12) for our problem.

The starting point of our investigation is the discussion of the formalism.
We introduce the $N$-flavor Nambu$-$Jona-Lasinio model in $D$-dimensional spacetime:
\begin{eqnarray}
{\cal L}(x) &=& \bar{\psi}(x)i\gamma^{\nu}(x)\nabla_{\nu}\psi(x)+\frac{\lambda}{2N}[(\bar{\psi}(x)\psi(x))^{2}+(\bar{\psi}(x)i\gamma_{5}\psi(x))^{2}] . 
\end{eqnarray}
Here, $\psi$ is the Dirac field, and $\lambda$ is the coupling constant of 
chiral invariant four-body contact interaction. 
The gamma-matrices $\gamma_{\mu}$, the metric $g_{\mu\nu}$ and the vielbein $e^{\mu}_{\hat{m}}$ are 
determined by the following relations~[13]:
\begin{eqnarray}
& & \{\gamma_{\mu}(x),\gamma_{\nu}(x)\}=2g_{\mu\nu}(x), \quad \{\gamma_{\hat{m}},\gamma_{\hat{n}}\}=2\eta_{\hat{m}\hat{n}}, \quad \eta_{\hat{m}\hat{n}}={\rm diag}(1,-1,-1,\cdots,-1), \nonumber \\
& & g_{\mu\nu}g^{\nu\rho}=\delta^{\rho}_{\mu}, \quad g^{\mu\nu}(x)=e^{\mu}_{\hat{m}}(x)e^{\nu \hat{m}}(x), \quad \gamma_{\mu}(x)=e^{\hat{m}}_{\mu}(x)\gamma_{\hat{m}}. 
\end{eqnarray}
Here, $\hat{m}$ refers to the flat tangent space defined by the vielbein at spacetime point $x$.
The definitions of covariant derivative $\nabla_{\nu}$ and 
spin connection $A^{\hat{m}\hat{n}}_{\nu}$ are given as follows~[13]:
\begin{eqnarray}
& & \nabla_{\nu}\equiv\partial_{\nu}-\frac{i}{2}A^{\hat{m}\hat{n}}_{\nu}\sigma_{\hat{m}\hat{n}},  \quad  \sigma_{\hat{m}\hat{n}}\equiv\frac{i}{4}[\gamma_{\hat{m}},\gamma_{\hat{n}}],    \nonumber \\
& & A^{\hat{m}\hat{n}}_{\mu}\equiv\frac{1}{2}e^{\hat{m}\lambda}e^{\hat{n}\rho}[C_{\lambda\rho\mu}-C_{\rho\lambda\mu}-C_{\mu\lambda\rho}], \quad C_{\lambda\rho\mu}\equiv e^{\hat{m}}_{\lambda}[\partial_{\rho}e_{\hat{m}\mu}-\partial_{\mu}e_{\hat{m}\rho}]. 
\end{eqnarray}  
The Rindler coordinates $(\eta,\xi,\bmx_{\perp})$ are related to the Minkowski coordinates
$(x_{0},x_{1},\bmx_{\perp})$ by the following coordinate transformation:
$x_{0}=\xi\sinh\eta, x_{1}=\xi\cosh\eta$. 
Under the transformation, the Minkowski spacetime is devided into two space-like wedges:
The right Rindler wedge,
$R_{+}=\{ x|x^{1}>|x^{0}|\} ( 0<\xi<+\infty, -\infty<\eta<+\infty )$,
and the left Rindler wedge,
$R_{-}=\{ x|x^{1}<-|x^{0}|\} ( -\infty<\xi<0, -\infty<\eta<+\infty )$.
$\eta$ is the time variable in Rindler coordinates.
$\xi=0$ corresponds to the event horizon of the Rindler spacetime.
These two wedges are causally disconnected with each other.
Hereafter, we concentrate on examining our problem in the right wedge.
The world line of the observer in Rindler coordinates is given as
\begin{eqnarray}
\eta(\tau)=a\tau, \quad \xi(\tau)=a^{-1}, \quad \bmx_{\perp}(\tau)={\rm const.},
\end{eqnarray}
where, $\tau$ is the proper time of an observer, and $a$ is a constant of acceleration.
The metric is chosen as $g_{\mu\nu}={\rm diag}(\xi^{2},-1,-1,\cdots,-1)$.
The line element of Rindler coordinates becomes
\begin{eqnarray}
ds^{2} &=& g_{\mu\nu}(x)dx^{\mu}dx^{\nu} =\xi^{2}d\eta^{2}-d\xi^{2}-d\bmx^{2}_{\perp}.
\end{eqnarray}
One obtains the gamma matrices in Rindler coordinates from the definition given in (2):
\begin{eqnarray}
\gamma^{0}(x)=\frac{1}{\xi}\gamma^{\hat{0}}, \quad \gamma^{1}(x)=\gamma^{\hat{1}}, \quad \gamma^{2}(x)=\gamma^{\hat{2}}, \quad\cdots,\quad \gamma^{D-1}(x)=\gamma^{\hat{D-1}}. 
\end{eqnarray}
Here, the $\gamma^{\hat{m}}, (m=0,1,2,\cdots,D-1)$ are the usual Dirac gamma-matrices 
of Minkowski spacetime. 
After computing the spin connection by the definition given in (3), 
one finds the components of the covariant derivatives in Rindler coordinates: 
\begin{eqnarray}
\nabla_{0}=\partial_{\eta}+\frac{1}{2}\gamma_{\hat{0}}\gamma_{\hat{1}}, \quad \nabla_{1}=\partial_{\xi}, \quad \nabla_{2}=\partial_{2}, \quad\cdots,\quad \nabla_{D-1}=\partial_{D-1}.  
\end{eqnarray}

Next, we introduce the following auxiliary fields:
\begin{eqnarray}
\sigma(x)\equiv-\frac{\lambda}{N}\bar{\psi}(x)\psi(x), \quad \pi(x)\equiv-\frac{\lambda}{N}\bar{\psi}(x)i\gamma_{5}\psi(x).
\end{eqnarray}
Then we obtain the partition function in our problem:
\begin{eqnarray}
Z &=& \int{\cal D}\bar{\psi}{\cal D}\psi{\cal D}\sigma{\cal D}\pi  \exp \Bigl\{ i N\int d^{D}x\sqrt{-g}[ -\frac{\sigma^{2}+\pi^{2}}{2\lambda} + \bar{\psi}(i\gamma^{\nu}(x)\nabla_{\nu}-\sigma-i\gamma_{5}\pi)\psi] \Bigr\} \nonumber \\
 &=& \int{\cal D}\sigma{\cal D}\pi \exp \Bigl\{ iN\int d^{D}x \sqrt{-g}[-\frac{\sigma^{2}+\pi^{2}}{2\lambda}] -i\ln{\rm Det}(i\gamma^{\nu}(x)\nabla_{\nu}-\sigma-i\gamma_{5}\pi) \Bigr\}.
\end{eqnarray}
Employing the large-$N$ expansion, the effective action integral in the leading order becomes
\begin{eqnarray}
S_{eff} &=& \int d^{D}x \sqrt{-g}[-\frac{\sigma^{2}+\pi^{2}}{2\lambda}] -i\ln{\rm Det}(i\gamma^{\nu}(x)\nabla_{\nu}-\sigma-i\gamma_{5}\pi).
\end{eqnarray}
Hereafter, we will omit the $\pi$ field because it is not needed for our purposes. 
Using the following relation
\begin{eqnarray}
\ln{\rm Det}(i\gamma^{\nu}\nabla_{\nu}-\sigma) &=& {\rm tr}\int d^{D}x \ln (i\gamma^{\nu}\nabla_{\nu}-\sigma) \nonumber \\
&=& -{\rm tr}\int d^{D}x\sqrt{-g}\int^{\sigma}_{0}ds S(x,x;s),
\end{eqnarray}
one obtains the effective action in the right wedge:
\begin{eqnarray}
S^{R^{(+)}}_{eff} &=& \int d^{D}x \sqrt{-g}(-\frac{\sigma^{2}}{2\lambda})+i{\rm tr}\int d^{D}x\sqrt{-g}\int^{\sigma}_{0}ds S(x,x;s).
\end{eqnarray}
Here we have introduced the Green's functions $S(x,y;s)$ 
and $G(x,y;s)$ satisfing the following equations:
\begin{eqnarray}
(i\gamma^{\nu}\nabla_{\nu}-s)S(x,y;s) &=& \frac{1}{\sqrt{-g}}\delta^{D}(x,y), \\
(i\gamma^{\nu}\nabla_{\nu}+s)G(x,y;s) &=& S(x,y;s), \\
(-\gamma^{\nu}\gamma^{\mu}\nabla_{\nu}\nabla_{\mu}-s^{2})G(x,y;s) &=& \frac{1}{\sqrt{-g}}\delta^{D}(x,y).
\end{eqnarray}
The Fourier transform of the Green's function $G$ is
\begin{eqnarray}
G(x,y;s) &=& G(\eta_{1}-\eta_{2},\xi_{1},\xi_{2},\bmx_{\perp}-\bmy_{\perp};s) \nonumber \\
&=& \int\frac{dk_{0}}{2\pi}\int\frac{d\bmk^{D-2}_{\perp}}{(2\pi)^{D-2}}e^{-ik_{0}(\eta_{1}-\eta_{2})+i\bmk_{\perp}\cdot(\bmx_{\perp}-\bmy_{\perp})}G(k_{0},\xi_{1},\xi_{2},\bmk_{\perp};s).
\end{eqnarray}

Next, we change our theory to the Euclidean formalism 
to incorporate with the thermal effect of acceleration~[8,9,14].
The Euclidean Rindler spacetime has a singularity at $\xi=0$.
To avoid it, we have to choose the period of the imaginary time as $2\pi$~[14,15].
The Euclidean formalism in Rindler coordinates with a definite period $\beta=2\pi$ of imaginary time 
coincides with the finite-temperature Matsubara formalism.
The Matsubara formalism is obtained by the following substitutions in our theory~[16]:
\begin{eqnarray}
\int\frac{dk_{0}}{2\pi}\to\sum_{n}\frac{1}{\beta}, &\quad& k_{0}\to i\omega_{n}, \nonumber \\
iS(k_{0},\xi_{1},\xi_{2},\bmk_{\perp};s)\to -{\cal S}(\omega_{n},\xi_{1},\xi_{2},\bmk_{\perp};s), &\quad& iG(k_{0},\xi_{1},\xi_{2},\bmk_{\perp};s)\to -{\cal G}(\omega_{n},\xi_{1},\xi_{2},\bmk_{\perp};s),  \nonumber \\
& & 
\end{eqnarray}
where, $\omega_{n}$ is the fermion discrete frequency defined 
as $\omega_{n}=(2n+1)\pi/\beta$ ($n=0, \pm 1, \pm 2, \cdots$). 
For the Green's functions, we use the abbreviations:
\begin{eqnarray}
\tilde{\cal S}(\xi,\xi';s) \equiv {\cal S}(\omega_{n},\xi,\xi',\bmk_{\perp};s), \quad \tilde{\cal G}(\xi,\xi';s) \equiv {\cal G}(\omega_{n},\xi,\xi',\bmk_{\perp};s).  
\end{eqnarray}
Some manipulations derive the following differential equation for $\tilde{\cal G}(\xi,\xi';s)$:
\begin{eqnarray}
\Bigl\{(\partial_{\xi}+\frac{1}{2\xi})(\partial_{\xi}+\frac{1}{2\xi})-(\bmk^{2}_{\perp}+s^{2}+\frac{\omega^{2}_{n}}{\xi^{2}})-\gamma_{\hat{0}}\gamma_{\hat{1}}\frac{\omega_{n}}{\xi^{2}} \Bigr\} \tilde{\cal G}(\xi,\xi';s) = \frac{1}{\xi}\delta(\xi,\xi').  
\end{eqnarray}
In Eq.(19), we divide the differential operator as follows:
\begin{eqnarray}
\hat{Q} \equiv \frac{d^{2}}{d\xi^{2}}+\frac{1}{\xi}\frac{d}{d\xi}-(\bmk^{2}_{\perp}+s^{2}+\frac{\omega^{2}_{n}+\frac{1}{4}}{\xi^{2}}), \quad \hat{R} \equiv -\frac{\omega_{n}}{\xi^{2}}.
\end{eqnarray}
We introduce the projection operators in the gamma-matrix space defined by
\begin{eqnarray}
P_{+}\equiv\frac{1}{2}(1+\gamma_{\hat{0}}\gamma_{\hat{1}}), \quad P_{-}\equiv\frac{1}{2}(1-\gamma_{\hat{0}}\gamma_{\hat{1}}), \quad \tilde{\cal G}=\tilde{\cal G}_{+}P_{+}+\tilde{\cal G}_{-}P_{-}.
\end{eqnarray}
Here, $P_{+}+P_{-}=1$, $P_{+}P_{+}=P_{+}$, $P_{-}P_{-}=P_{-}$, and $P_{+}P_{-}=P_{-}P_{+}=0$ are satisfied.
Thus, Eq. (19) can be written as follows: 
\begin{eqnarray}
(\hat{Q}+\gamma_{\hat{0}}\gamma_{\hat{1}}\hat{R})\tilde{\cal G}(\xi,\xi';s) &=& ((\hat{Q}+\hat{R})P_{+}+(\hat{Q}-\hat{R})P_{-})\tilde{\cal G}(\xi,\xi';s)= \frac{1}{\xi}\delta(\xi,\xi'). 
\end{eqnarray}
Therefore, Eq.(22) is decoupled into the following two equations:
\begin{eqnarray}
(\hat{Q}+\hat{R})\tilde{\cal G}_{+}(\xi,\xi';s)=\frac{1}{\xi}\delta(\xi,\xi'), \quad (\hat{Q}-\hat{R})\tilde{\cal G}_{-}(\xi,\xi';s)=\frac{1}{\xi}\delta(\xi,\xi'). 
\end{eqnarray}
The eigenfunctions of the operators $\hat{Q}\pm\hat{R}$ are given as 
\begin{eqnarray}
\Psi^{+}_{\Omega}(\xi) \equiv \frac{\sqrt{(-2i\Omega-1)\cosh\pi\Omega}}{\pi}K_{i\Omega+\frac{1}{2}}(\alpha\xi), &\quad& \Psi^{-}_{\Omega}(\xi) \equiv \frac{\sqrt{(+2i\Omega-1)\cosh\pi\Omega}}{\pi} K_{i\Omega-\frac{1}{2}}(\alpha\xi). \nonumber \\
& &  
\end{eqnarray}
Here, $\alpha = \sqrt{\bmk^{2}_{\perp}+s^{2}}$, and $K_{\mu}(x)$ is the modified Bessel function. 
The orthonormal condition for $\Psi^{\pm}_{\Omega}$ is~[17,18]
\begin{eqnarray}
\int^{\infty}_{0}\frac{d\xi}{\xi}\Psi^{\pm}_{\Omega}(\xi)\Psi^{\pm}_{\Omega'}(\xi) = \int^{\infty}_{0}\frac{d\xi}{\xi}\frac{(\mp2i\Omega-1)\cosh\pi\Omega}{\pi^{2}}K_{i\Omega\pm\frac{1}{2}}(\alpha\xi)K_{i\Omega'\pm\frac{1}{2}}(\alpha\xi) = \delta(\Omega,\Omega'). 
\end{eqnarray}
Expanding $\tilde{\cal G}_{\pm}$ by the eigenfunctions $\Psi^{\pm}_{\Omega}$, 
and using the orthonormal relation, $\tilde{\cal G}_{\pm}$ are obtained in the following form:
\begin{eqnarray}
\tilde{\cal G}_{\pm}(\xi,\xi';s) &=& -\int^{\infty}_{0} d\Omega \frac{\Psi^{\pm}_{\Omega}(\xi)\Psi^{\pm}_{\Omega}(\xi')}{(\omega_{n}\pm\frac{1}{2})^{2}-(i\Omega\pm\frac{1}{2})^{2}} \nonumber \\
&=& \int^{\infty}_{0}d\Omega\frac{(\mp 2i\Omega-1)\cosh\pi\Omega}{\pi^{2}}\frac{K_{i\Omega\pm\frac{1}{2}}(\alpha\xi)K_{i\Omega\pm\frac{1}{2}}(\alpha\xi')}{(i\omega_{n}-\Omega\pm i)(i\omega_{n}+\Omega)}.
\end{eqnarray}

The position of the uniformly accelerated observer in Rindler coordinates is given by (4).
To obtain the value of the effective potential at this position,
we have to change the variables as $\eta\to a\tau$ and $\xi\to a^{-1}$.
The definition of the effective potential is 
\begin{eqnarray}
V_{eff} &=& -\frac{S_{eff}}{\int d^{D}x \sqrt{-g}}.
\end{eqnarray}
Hence in our case
\begin{eqnarray}
V^{R^{(+)}}_{eff}(a,\sigma) &=& \frac{\sigma^{2}}{2\lambda} + {\rm tr}\sum_{n}\frac{1}{\beta}\int^{\sigma}_{0}ds\int\frac{d^{D-2}\bmk_{\perp}}{(2\pi)^{D-2}}  \tilde{\cal S}(a^{-1},a^{-1};s).
\end{eqnarray}
$V^{R^{(+)}}_{eff}(a,\sigma)$ is a local, intensive quantity.
The effective potential is normalized as $V^{R^{(+)}}_{eff}(a,\sigma=0)=0$.
The gap equation corresponds to the stationary condition:
\begin{eqnarray}
0 &=& \frac{\partial V_{eff}}{\partial \sigma}.
\end{eqnarray}
Therefore, the self-consistency condition is derived as
\begin{eqnarray}
\sigma &=& -\lambda {\rm tr}\sum_{n}\frac{1}{\beta} \int\frac{d^{D-2}\bmk_{\perp}}{(2\pi)^{D-2}} \tilde{\cal S}(a^{-1},a^{-1};\sigma).
\end{eqnarray}
From the following relation
\begin{eqnarray}
\Bigl(\frac{i}{\xi}\gamma_{\hat{0}}\omega_{n}+i\gamma_{\hat{1}}(\partial_{\xi}+\frac{1}{2\xi})-\vec{\gamma}_{\perp}\cdot\bmk_{\perp}+s \Bigr)\tilde{\cal G}(\xi,\xi';s) &=& \tilde{\cal S}(\xi,\xi';s), 
\end{eqnarray}
the gap equation is found to be
\begin{eqnarray}
\sigma &=& -\lambda{\rm tr}\sum_{n}\frac{1}{\beta}\int\frac{d^{D-2}\bmk_{\perp}}{(2\pi)^{D-2}}\sigma\tilde{\cal G}(a^{-1},a^{-1};\sigma)   \nonumber \\
&=& -\lambda\sigma{\rm tr}\sum_{n}\frac{1}{2\pi}\int\frac{d^{D-2}\bmk_{\perp}}{(2\pi)^{D-2}}\int^{\infty}_{0}d\Omega    \nonumber \\
& & \times \Bigl\{ P_{+}\frac{(-2i\Omega-1)\cosh\pi\Omega}{\pi^{2}}(K_{i\Omega+\frac{1}{2}}(\alpha a^{-1}))^{2}\frac{1}{(i\omega_{n}-\Omega+i)(i\omega_{n}+\Omega)}  \nonumber \\
& & \qquad +P_{-}\frac{(2i\Omega-1)\cosh\pi\Omega}{\pi^{2}}(K_{i\Omega-\frac{1}{2}}(\alpha a^{-1}))^{2}\frac{1}{(i\omega_{n}-\Omega-i)(i\omega_{n}+\Omega)} \Bigr\}   \nonumber \\
&=& -\frac{2i\lambda\sigma}{a} \int\frac{d^{D-2}\bmk_{\perp}}{(2\pi)^{D-2}}\int^{\infty}_{0}d\omega \frac{\sinh\frac{\pi}{a}\omega}{\pi^{2}}\Bigl\{(K_{i\frac{\omega}{a}+\frac{1}{2}}(\alpha a^{-1}))^{2}-(K_{i\frac{\omega}{a}-\frac{1}{2}}(\alpha a^{-1}))^{2} \Bigr\}, 
\end{eqnarray}
where, $\alpha=\sqrt{\bmk^{2}_{\perp}+\sigma^{2}}$.
To obtain the final expression in Eq. (32), the frequency summation was performed,
and the integration variable was changed as $\Omega\to\omega/a$.

By using the result given above, we first examine 
the equation for the determination of critical acceleration for symmetry restoration.
Hereafter, we restrict ourselves to the four-dimensional case $D=4$.
Setting $\sigma=0$ in Eq. (32) of the case of nontrivial solution, one finds
\begin{eqnarray}
1=-\frac{i\lambda}{\pi^{3}a} \int^{\infty}_{0}kdk \int^{\infty}_{0}d\omega \sinh\frac{\pi}{a}\omega\Bigl\{(K_{i\frac{\omega}{a}+\frac{1}{2}}(\frac{k}{a}))^{2}-(K_{i\frac{\omega}{a}-\frac{1}{2}}(\frac{k}{a}))^{2} \Bigr\}. 
\end{eqnarray}
Integration in $k$ is performed by making use of the following formula~[17,18]:
\begin{eqnarray}
& & \int^{\infty}_{0}dxK_{m}(\alpha x)K_{n}(\alpha x)x^{l} = 2^{l-2}\alpha^{-l-1}\Gamma(l+1)^{-1} \nonumber \\
& & \times \Gamma\Bigl(\frac{l+m+n+1}{2}\Bigr)\Gamma\Bigl(\frac{l+m-n+1}{2}\Bigr)\Gamma\Bigl(\frac{l-m+n+1}{2}\Bigr)\Gamma\Bigl(\frac{l-m-n+1}{2}\Bigr). 
\end{eqnarray}
( Valid when $\Re (l-m-n) > -1$. ) Then we obtain
\begin{eqnarray}
1 &=& \frac{\lambda}{\pi^{2}}\int^{\infty}_{0}\omega d\omega\tanh\frac{\pi}{a}\omega = \frac{\lambda}{\pi^{2}}\int^{\infty}_{0} d\omega\Bigl\{ \omega-\frac{2\omega}{\exp\frac{2\pi\omega}{a}+1}\Bigr\}.
\end{eqnarray} 
The integration in Eq. (35) yields the equation for the determination of critical acceleration $a_{c}$:
\begin{eqnarray}
1 &=& \frac{\lambda\Lambda^{2}}{2\pi^{2}}-\frac{\lambda a^{2}_{c}}{24\pi^{2}}.
\end{eqnarray}
To obtain the above expression, we have simply used a cutoff $\Lambda$ for the regularization.
Because $a_{c}$ is positive, the critical Hawking-Unruh temperature becomes 
\begin{eqnarray}
T^{(c)}_{U} &\equiv& \frac{a_{c}}{2\pi} = \sqrt{\frac{3}{\pi^{2}}\Lambda^{2}-\frac{6}{\lambda}}.
\end{eqnarray}
In order to compare with the usual thermal case, let us refer to the result in Minkowski spacetime~[19].
The gap equation in finite-temperature case is 
\begin{eqnarray}
1 &=& \frac{\lambda}{2\pi^{2}}\Bigl\{ \Lambda\sqrt{\Lambda^{2}+\sigma^{2}}-\sigma^{2}\ln\frac{\Lambda+\sqrt{\Lambda^{2}+\sigma^{2}}}{\sigma}\Bigr\}-\frac{2\lambda}{\pi^{2}}\int^{\infty}_{0}dk \frac{k^{2}}{\sqrt{k^{2}+\sigma^{2}}}\frac{1}{e^{\sqrt{k^{2}+\sigma^{2}}/T}+1}.
\end{eqnarray}
Hence, at $\sigma=0$, our equation (36) coincides with the case of usual finite-temperature gap equation 
in Minkowski spacetime.
The critical Hawking-Unruh temperature depends on $\Lambda$ and $\lambda$
in the same way as the case of the temperature of thermal restoration in Minkowski spacetime.
When $\Lambda$ and $\lambda$ become large, $T^{(c)}_{U}$ increases. 
In the above expression, $\frac{3\Lambda^{2}}{\pi^{2}}-\frac{6}{\lambda}>0$ has to be satisfied.
We assume this condition indicates the existence of a critical coupling $\lambda_{c}$.
Thus we arrive at the familiar expression:
\begin{eqnarray}
\lambda_{c} &=& \frac{2\pi^{2}}{\Lambda^{2}}.
\end{eqnarray}

Now, we give a rough estimation for $a_{c}$. 
The Unruh temperature is given by 
$T_{U}=\hbar a/(2\pi k_{B}c)=a/(2.5\times10^{22}({\rm cm\cdot s^{-2}})){\rm K}$ 
( $c$ is the velocity of light, and $k_{B}$ is the Boltzmann constant ).
Here we consider the case of quark, assume $a_{c}\sim \Lambda\times 10^{-1}$, 
and choose $\Lambda=\Lambda_{QCD}\sim 1 {\rm GeV}$~[19].
Because $1{\rm GeV}=1.2\times10^{13}{\rm K}$, one has 
\begin{eqnarray}
a_{c} &\sim& 3\times10^{34}{\rm cm\cdot s^{-2}}.
\end{eqnarray}
This value belongs to an extremely high acceleration regime.
The observer almost have to be accelerated to the order of this value 
to observe the chiral symmetry restoration of a quark.
A massive Dirac particle will be observed as a massless particle above $a_{c}$.

Next, we examine the behaviour of our gap equation at $a\to a_{c}$.
The gap equation (32) is
\begin{eqnarray}
1 &=& -i\frac{\lambda a^{2}}{\pi^{3}}\int^{\Lambda/a}_{0}dx\sinh\pi x\int^{\infty}_{\sigma/a}ydy \Bigl\{ (K_{ix+\frac{1}{2}}(y))^{2}-(K_{ix-\frac{1}{2}}(y))^{2}   \Bigr\}  \nonumber \\
&=& -i\lambda\frac{\sigma^{2}}{2\pi^{3}}\int^{\Lambda/a}_{0}dx\sinh\pi x \nonumber \\
& & \times\Bigl\{ (K_{ix-\frac{1}{2}}(\frac{\sigma}{a}))^{2}-(K_{ix+\frac{1}{2}}(\frac{\sigma}{a}))^{2}+K_{ix-\frac{1}{2}}(\frac{\sigma}{a})K_{ix+\frac{3}{2}}(\frac{\sigma}{a})-K_{ix+\frac{1}{2}}(\frac{\sigma}{a})K_{ix-\frac{3}{2}}(\frac{\sigma}{a}) \Bigr\}.
\end{eqnarray}
When $a\to a_{c}$, $\sigma/a\to 0$ may occur. 
This condition corresponds to the case when the argument $z$ of the Bessel function $K_{\mu}(z)$ tends to zero.
By using the forms of the Bessel functions at $\sigma/a\to 0$, we find:
\begin{eqnarray}
1 &=& \frac{\lambda}{\pi^{2}}a^{2}\int^{\Lambda/a}_{0}x dx\tanh\pi x -i\frac{\lambda a}{4\pi}\sigma\int^{\infty}_{-\infty}dx\frac{\sinh\pi x}{\cosh^{2}\pi x}\frac{(\sigma/2a)^{2ix}}{\Gamma(ix+\frac{1}{2})^{2}}  \nonumber \\
&=& \frac{\lambda\Lambda^{2}}{2\pi^{2}}-\frac{\lambda a^{2}}{24\pi^{2}} -\frac{2\lambda a^{2}}{\pi}\sum^{\infty}_{n=1}(\frac{\sigma}{2ia})^{2n}\frac{1}{(n-1)!}\Bigl(\ln\frac{\sigma}{2a}-\psi(n) \Bigr).
\end{eqnarray}
Here, $\psi(n)$ is the digamma function: $\psi(n)\equiv\sum^{n-1}_{j=1}\frac{1}{j}-\gamma$ ( $\gamma$ is the Euler constant; 0.577... ).
To obtain the final expression, we have performed an appropriate contour integration.
Picking the largest contribution under the condition $\sigma\ll 2a$, we get
\begin{eqnarray}
1 &=& \frac{\lambda\Lambda^{2}}{2\pi^{2}}-\frac{\lambda a^{2}}{24\pi^{2}} -\frac{\lambda}{2\pi}\sigma^{2}\ln (e^{-\gamma}\frac{2a}{\sigma}).
\end{eqnarray}
The combination of this equation with Eq. (36) gives
\begin{eqnarray}
a^{2}_{c}-a^{2} &=& 12\pi\sigma^{2}\ln(e^{-\gamma}\frac{2a}{\sigma}).
\end{eqnarray}
Taking the derivative with respect to $a$, we have
\begin{eqnarray}
\frac{d\sigma}{da} \approx -\frac{a}{12\pi\sigma\ln(e^{-\gamma}2a/\sigma)} = \sigma\frac{a}{a^{2}-a^{2}_{c}}  < 0,
\end{eqnarray}
under the condition $\sigma \ll 2a$. The solution of this differential equation is
\begin{eqnarray}
\sigma &=& e^{C_{1}}\sqrt{a^{2}_{c}-a^{2}}. 
\end{eqnarray}
$C_{1}$ is an integration constant.
In the vicinity of the phase transition, $\sigma$ depends on $a$ in the way of Eq. (46).
Because the gap equation is equivalent to $\partial V^{R^{(+)}}_{eff}/\partial\sigma=0$, 
we arrive at the following expression for the effective potential from Eq.(42):
\begin{eqnarray}
V^{R^{(+)}}_{eff}(a,\sigma) &=& \int^{\sigma}_{0}sds\Bigl\{ \frac{1}{\lambda}-\frac{\Lambda^{2}}{2\pi^{2}}+\frac{a^{2}}{24\pi^{2}} +\frac{2a^{2}}{\pi}\sum^{\infty}_{n=1}(-\frac{s^{2}}{4a^{2}})^{n}\frac{1}{(n-1)!}(\frac{1}{2}\ln\frac{s^{2}}{4a^{2}}-\psi(n))  \Bigr\} \nonumber \\
&=& -\frac{1}{48\pi^{2}}(a^{2}_{c}-a^{2})\sigma^{2} +\frac{1}{8\pi}\Bigl(\ln\frac{2a}{\sigma}+\frac{1}{4}-\gamma \Bigr)\sigma^{4} + \cdots  \nonumber \\
&=& -\frac{1}{96\pi^{2}}(a^{2}_{c}-a^{2})\sigma^{2} +\frac{1}{32\pi}\sigma^{4} + \cdots.
\end{eqnarray}
To obtain the final expression, we have used Eqs.(36) and (44).
This result is a kind of Ginzburg-Landau-type energy functional.
The effective potential (47) clearly shows 
the phenomenon of broken symmetry by its wine bottle-like shape 
under the variation with respect to $\sigma$ ( in the phase mode ).
Taking the derivative with respect to $a$ in Eq. (47), 
we find that this potential gives the second-order phase transition.
From Eq. (46), we find the critical exponent of the order parameter $\sigma$ is $1/2$,
and this coincides with the case of the Landau theory of second-order phase transition.

In summary, we have obtained several important results: 
(1) The critical Hawking-Unruh temperature of the chiral mass for a uniformly accelerated observer 
is given as $T^{(c)}_{U}=a_{c}/2\pi=\sqrt{3\Lambda^{2}/\pi^{2}-6/\lambda}$, 
which coincides with the case of the thermal restoration of the dynamical chiral symmetry breaking 
of the Nambu$-$Jona-Lasinio model in Minkowski spacetime. 
(2) Based on the thermalization theorem, we have found that 
the effect of the Unruh temperature will cause the restoration of the broken chiral symmetry. 
A point-like massive Dirac particle located at the position of the accelerated observer
can give the chiral symmetry restoration.
By the examination in the vicinity of the transition, we have found that 
the effective potential ( free energy density ) is written as a Ginzburg-Landau functional,
and it gives a second order phase transition.

Finally, we would like to make some comments on several issues and possible extensions of this work.
The Rindler metric can be regarded as an approximation of the metric of 
a large mass Schwarzschild black hole near the outside of the event horizon~[14].
Thus the physics in Rinlder wedge provides us with a simplified model 
to study the situation around a Schwarzschild black hole. 
Investigations of the phenomena of dynamical symmetry breaking in such 
a large mass black hole is also an interesting problem.
It is also important for us to study the finite-density case, by using our model.
In this work, we have only considered the case of scalar fermion-antifermion condensate
without a chemical potential.
We can extend our theory to consider the other cases of symmetries,
such as vector, tensor and so forth.
Moreover, a treatment of Majorana-type mass with the fermion-fermion condensate $\langle\psi\psi\rangle$
is also interesting. This issue might be related to the problem of neutrino Majorana mass.
The meson-diquark bosonization of the Nambu$-$Jona-Lasinio model~[20] in Rindler coordinates
is possible. In such a case, to take into account the Pauli-G\"{u}rsey symmetry
would give us a useful point of view~[21]. If the Pauli-G\"{u}rsey symmetry is realized
in the Lagrangian, there is a rotational symmetry between meson and diquark. 
We speculate that the phenomena we have found in this paper will also occur in  a Majorana-type mass.

\acknowledgments

The author would like to express his gratitude sincerely to 
Professors Yoichiro Nambu and Kazumi Okuyama, 
for their enlightening discussions, comments and encouragements.


\begin{thebibliography}{999} 



\bibitem{hawking1} 
S. W. Hawking, Commun. Math. Phys. {\bf 43}, 199 (1975), 
S. A. Fulling, Phys. Rev. {\bf D7}, 2850 (1973),
P. C. W. Davies, J. Phys. {\bf A8}, 609 (1975),
W. G. Unruh, Phys. Rev. {\bf D14}, 870 (1976).
\bibitem{wald1}
W. G. Unruh and R. M. Wald, Phys. Rev. {\bf D25}, 942 (1982),
ibid, {\bf D27}, 2271 (1983), ibid, {\bf D29}, 1047 (1984),
W. G. Unruh, ibid. {\bf D46}, 3271 (1992).
\bibitem{troost}
W. Troost and H. Van Dam, Nucl. Phys. {\bf B152}, 442 (1979).
\bibitem{cduff}
S. M. Christensen and M. J. Duff, Nucl. Phys. {\bf B146}, 11 (1978).
\bibitem{dowker}
J. S. Dowker, J. Phys. {\bf A10}, 115 (1977), Phys. Rev. {\bf D18}, 1856 (1978).
\bibitem{netsu-green}
G. W. Gibbons and M. J. Perry, Proc. Roy. Soc. Lond. {\bf A358}, 467 (1978).
\bibitem{sewell}
G. L. Sewell, Ann. Phys. {\bf 141}, 201 (1982).
\bibitem{unrwei} 
W. G. Unruh and N. Weiss, Phys. Rev. {\bf D29}, 1656 (1984).
\bibitem{lee}
T. D. Lee, Nucl. Phys. {\bf B264}, 437 (1986), 
R. Friedberg, T. D. Lee and Y. Pang, ibid. {\bf B276}, 549 (1986).
\bibitem{takagi} 
S. Takagi, Prog. Theor. Phys. Suppl. {\bf 88}, 1 (1986).
\bibitem{nambu2}
Y. Nambu and G. Jona-Lasinio, Phys. Rev. {\bf 122}, 345 (1961), ibid. {\bf 124}, 246 (1961).
\bibitem{inagaki}
Extensive studies on the dynamical symmetry breaking in curved spacetime are found in 
the following review:
T. Inagaki, T. Muta and S. D. Odintsov, Prog. Theor. Phys. Suppl. {\bf 127}, 93 (1997).
\bibitem{vierbein}
For the definitions of vielbein and spin connection for Dirac field in curved background, see:
D. R. Brill and J. A. Wheeler, Rev. Mod. Phys. {\bf 29}, 465 (1957),
L. Parker and D. J. Toms, Phys. Rev. {\bf D29}, 1584 (1984).
\bibitem{euclid}
The method of Euclidean path integral and effective action in Rindler spacetime:
G. Cognola, K. Kirsten and L. Vanzo, Phys. Rev. {\bf D49}, 1029 (1994),
D. Kabat and M. J. Strassler, Phys. Lett. {\bf B329}, 46 (1994),
D. Kabat, Nucl. Phys. {\bf B453}, 281 (1995),
R. Emparan, Phys. Rev. {\bf D51}, 5716 (1995),
A. A. Bytsenko, G. Cognola and S. Zerbini, Nucl. Phys. {\bf B458}, 267 (1996),
S. Zerbini, G. Cognola and L. Vanzo, Phys. Rev. {\bf D54}, 2699 (1996),
D. Iellici and V. Moretti, Phys. Rev. {\bf D54}, 7459 (1996).
\bibitem{conical}
Discussions on the method to remove singularities in several spacetimes, see:
S. W. Hawking, Phys. Rev. {\bf D18}, 1714 (1978),
G. W. Gibbons and S. W. Hawking, Commun. Math. Phys. {\bf 66}, 291 (1979).
\bibitem{finitemp}
A. A. Abrikosov, L. P. Gor'kov and I. E. Dzyaloshinskii, 
{\it Methods of Quantum Field Theory in Statistical Physics} (Dover, New York, 1963),
J. I. Kapusta, {\it Finite-temperature Field Theory} (Cambridge University Press, Cambridge, 1989).
\bibitem{gfuncrind}
Green's functions in Rindler coordinates, 
for bosonic fields;
P. Candelas and D. Deutsch, Proc. Roy. Soc. Lond. {\bf A354} 79, (1977),
and for fermionic fields;
P. Candelas and D. Deutsch, Proc. Roy. Soc. Lond. {\bf A362} 251, (1978).
\bibitem{terashima}
H. Terashima, Phys. Rev. {\bf D60}, 084001 (1999).
\bibitem{klevansky}
S. P. Klevansky, Rev. Mod. Phys. {\bf 64}, 649 (1992).
\bibitem{bosonize}
For example: 
T. Eguchi and H. Sugawara, Phys. Rev. {\bf D10}, 4257 (1974),
T. Eguchi, Phys. Rev. {\bf D14}, 2755 (1976),
K. Kikkawa, Prog. Theor. Phys. {\bf 56}, 947 (1976),
D. Kahana and U. Vogl, Phys. Lett. {\bf B244}, 10 (1990),
D. Ebert, L. Kaschluhn and G. Kastelewicz, Phys. Lett. {\bf B264}, 420 (1991),
D. Ebert, Yu. L. Kalinovsky, L. M\"{u}nchow and M. K. Volkov, Int. J. Mod. Phys. {\bf A8}, 1295 (1993).
\bibitem{pgsymm}
W. Pauli, Nuovo Cim. {\bf 6}, 204 (1957), 
G. G\"{u}rsey, Nuovo Cim. {\bf 7}, 411 (1958).



\end{thebibliography}
\end{document}